\documentclass[twocolumn,english,aps,prl,superscriptaddress]{revtex4}
\usepackage[T1]{fontenc}
\usepackage{amsmath}
\usepackage{amssymb}
\usepackage{fancyhdr}     
\usepackage{amsfonts}
\usepackage[latin1]{inputenc}
\usepackage{graphicx,graphics,amssymb,times}
\usepackage{color}
\usepackage{natbib}
\usepackage{comment}
\usepackage{babel}

\makeatletter

\newcommand{\ket}[1]{| #1 \rangle}

\begin{document}

\title{Mutual Unbiasedness in Coarse-grained Continuous Variables}

\author{Daniel S. Tasca}
\email{dan.tasca@gmail.com}
\affiliation{Instituto de F\'isica, Universidade Federal Fluminense, Niter\'oi, RJ
24210-346, Brazil}

\author{Piero S\'{a}nchez}
\affiliation{Departamento de Ciencias, Secci\'on F\'isica, Pontificia Universidad
Cat\'olica del Per\'u, Apartado 1761, Lima, Peru}

\author{Stephen P. Walborn}
\affiliation{Instituto de F\'isica, Universidade Federal do Rio de Janeiro, Caixa
Postal 68528, Rio de Janeiro, RJ 21941-972, Brazil}

\author{\L ukasz Rudnicki}
\email{rudnicki@cft.edu.pl}
\selectlanguage{english}%
\affiliation{Max-Planck-Institut f{\"u}r die Physik des Lichts, Staudtstra{\ss}e 2, 91058
Erlangen, Germany}
\affiliation{Center for Theoretical Physics, Polish Academy of Sciences, Al.
Lotnik\'ow 32/46, 02-668 Warsaw, Poland}

\begin{abstract}
The notion of mutual unbiasedness for coarse-grained measurements of quantum continuous variable systems is considered.  It is shown that while the procedure of ``standard'' coarse graining breaks the mutual unbiasedness between conjugate variables, this desired feature can be theoretically established and experimentally observed in periodic coarse graining. We illustrate our results in an optics experiment implementing Fraunhofer diffraction through a periodic diffraction grating, finding excellent agreement with the derived theory. Our results are an important step in developing a formal connection between discrete and continuous variable quantum mechanics. 
\end{abstract}

\maketitle

\paragraph{Introduction.}
The ability to measure a system in an infinite number of non-commuting bases distinguishes the quantum world from classical physics.  Wave-particle duality and more generally the complementarity principle are directly rooted in this feature of quantum mechanics.  Though one can measure a quantum system in several distinct bases, uncertainty relations limit the amount of information that can be obtained. 
It is well known that projection onto an eigenstate of one basis reduces the information that can be obtained through or inferred about subsequent measurement in a different basis. The information is minimum for mutually unbiased bases (MUBs), for which all outcomes of the second measurement are equally likely, so that total uncertainty is always substantial (the sharpest uncertainty relations \cite{Sanchez93}) and most insensitive to input states \cite{Puchaa15}. MUBs play an important role in complementarity \cite{hall15}, quantum cryptography \cite{coles17} and quantum tomography \cite{perez11,Giovannini13}, are useful for certifying quantum randomness \cite{vallone14}, and for detecting quantum correlations such as entanglement \cite{spengler12,paul16,sauerwein17} and steering \cite{cavalcanti09,walborn11,tasca13,schneeloch13,schneeloch13b,Cavalcanti15,Horodecki15,zhu16}.   

Quantum information encoding in high-dimensional systems harbor the potential for efficient quantum cryptography \cite{Pirandola15,Krenn16,Vienna17} and interesting fundamental studies \cite{Dada11a,Potocek15}. A number of modern day implementations of high-dimensional quantum systems rely on continuous variables (CV) encoded in photonics systems. These CV \cite{Plenio16}  or hybrid \cite{Furusawa} platforms allow one to encode several bits per outcome. However, a typical measurement device does not register a continuous and infinite range of values, and it is thus necessary to consider discretized measurements. A most common approach is the selection of a finite set of transverse spatial modes labeled by discrete mode indexes \cite{Salakhutdinov12,Krenn14,Bobrov15,Restuccia16}, for which MUB measurements are attainable by the use of phase holograms \cite{Giovannini13}.
Free-space \cite{Neves07}, multi-core fibers \cite{Canas17} or on-chip \cite{Schaeff12} path encoding as well as time-bin \cite{Xie15} are also interesting techniques with potencial for high-dimensionality. These methods, despite being useful, discard a fraction of available modes and do not straightforwardly extend to the complementary (Fourier) domain of CVs. A different discretization procedure is the {\it coarse-graining} of the continuous degree of freedom itself \cite{rudnicki12a,rudnicki12b}. In this case, practical constraints such as finite detector resolution, or limited measurement time and sampling range, if not properly handled, can lead to false conclusions in tasks such as entanglement detection and cryptographic security \cite{tasca13, Ray13a,Ray13b}. 

The notion of quantum mechanical mutual unbiasedness seems rather well established. In particular, two orthonormal bases $\left|a_{i}\right\rangle $ and $\left|b_{j}\right\rangle $, $i,j=0,\ldots,d-1$, in a finite-dimensional Hilbert space (of dimension $d$) are mutually unbiased if and only if $\left|\left\langle a_{i}\left|b_{j}\right\rangle \right.\!\right|=1/\sqrt{d}$ for all $i,j$ (MUB condition) \cite{Durt10}. This definition (for finite $d$) can be extended to deal with--so called--mutually unbiased (non-projective \footnote{Mutually unbiased measurements are projective only in the limiting case when they describe the MUBs.}) measurements \cite{Kalev}. For continuous variables (CV), such as a conjugate pair formed by position and momentum, mutual unbiasedness is encoded in the relation 
$\left|\left\langle x\left|p\right\rangle \right.\!\right|=1/\sqrt{2\pi\hbar}$ \cite{Weigert08}. 
One might suspect that coarse graining preserves the original unbiasedness of continuous variables.  We shall explain below why this is not the case for standard coarse graining, and further construct a set of coarse-grained mutually unbiased measurements of finite cardinality. The latter property may in principle allow for a conceptual relationship between CV and finite dimensional quantum mechanics. 
We demonstrate our results in an optics experiment exploring the transverse position and momentum of a paraxial light field as conjugate CVs. The proposed coarse-graining is implemented and mutual unbiasedness is observed for measurement dimensionality up to $d=15$. Our coarse-graining model, in contrast to most of the discretization methods mentioned above, does not rely on the selection of a {\it subspace} of all available modes and delivers the MUBs in complementary domains. Thus, mutual unbiasedness in high-dimensional measurements is achieved without the assumption that operations \cite{Schlederer16,Babazadeh17} will not transfer the photon state out of the relevant subspace.

\paragraph{Unbiased coarse-grained measurements.} 
In the most basic scenario, one can describe experimental outcomes of coarse-grained position-momentum measurements by means of the projectors  \cite{rudnicki12a,rudnicki12b,schneeloch13,tasca13}  ($k,l\in\mathbb{Z}$):
\begin{equation}
A_{k}=\int_{k_{-}\Delta}^{k_{+}\Delta}\!\!\!dx\left|x\right\rangle \left\langle x\right|,\quad B_{l}=\int_{l_{-}\delta}^{l_{+}\delta}\!\!\!dp\left|p\right\rangle \left\langle p\right|,\label{projectors}
\end{equation}
with $j_{\pm}=j\pm1/2$. The two parameters $\Delta$ and $\delta$ are
the coarse-graining widths, which can be understood as resolutions of the detectors
used in an experiment. Looking at the explicit representations it
is quite easy to deduce that $\textrm{Tr}\left(A_{k}B_{l}\right)=\Delta\delta/2\pi\hbar$.
One however shall have in mind that mathematical underpinning of this
observation is highly nontrivial because the operators $A_{k}B_{l}$
are all of trace class, even though neither $A_{k}$ nor $B_{l}$
have this property. The fact that the overlap does not depend on
indices $k$ and $l$ suggests that the operators in (\ref{projectors}) are interrelated
in a special way --- so-called Accardi complementarity \cite{Accardi}. Note, however, that the constant-trace condition alone is not even enough to assure that original variables are connected by the Fourier transformation \cite{Cassinelli}. 
On the other hand, a quantum state localized in one coarse-grained
basis, for instance $\psi\left(x\right)=1/\sqrt{\Delta}$ for $\left|x\right|\leq\Delta/2$
(and $0$ elsewhere) that is covered by $A_{0}$, \textit{is not}
evenly spread with respect to the second one (here given by $\left\{ B_{l}\right\} $)
but instead decays like the \textit{sinc} function. 

In the last observation, we actually pointed out that the pair of
projective measurements (\ref{projectors}) does not meet the \emph{most
natural} definition of mutual unbiasedness in discrete settings that can be formulated
as follows. Given a pure state $\left|\Psi\right\rangle $ and two
sets of $d$ projective measurements, $\left\{ \Pi_{k}\right\} $
and $\left\{ \Omega_{l}\right\} $, we define the usual probabilities
$q_{k}\left(\Psi\right)=\left\langle \Psi\right|\Pi_{k}\left|\Psi\right\rangle $
and $p_{l}\left(\Psi\right)=\left\langle \Psi\right|\Omega_{l}\left|\Psi\right\rangle $.
The measurements are mutually unbiased if for all $\left|\Psi\right\rangle $
and all $k_{0},l_{0}=0,\ldots,d-1$:\begin{subequations}\label{MUBs}
\begin{equation}
q_{k}\left(\Psi\right)=\delta_{k_{0}k}\;\Longrightarrow\;p_{l}\left(\Psi\right)=d^{-1},\label{MUB1}
\end{equation}
\begin{equation}
p_{l}\left(\Psi\right)=\delta_{l_{0}l}\;\Longrightarrow\;q_{k}\left(\Psi\right)=d^{-1}.\label{MUB2}
\end{equation}
\end{subequations}Again in words, whenever the state is localized
in one set, it is evenly spread in the second one. The case with countably
infinite sets of projectors (like those in Eq. (\ref{projectors}),
which however do not fit into the definition) shall be understood
in the limit $d\rightarrow\infty$. Extension to genuinely continuous scenario would require subtle modifications of the definition; this case is however beyond our interest here. Whenever the pairs of projective
measurements are unitarily equivalent, a single requirement is sufficient.
Quite obviously, this definition correctly reproduces the MUB condition.

As standard coarse graining \eqref{projectors} does not satisfy the definition \eqref{MUBs}, we consider now another type of coarse graining.  In general, one can define projectors 
\begin{subequations}\label{Proj}
\begin{equation}
\Pi_{k}=\int_{\mathbb{R}}dxM_{k}\left(x-x_{\textrm{cen}};T_{x}\right)\left|x\right\rangle \left\langle x\right|,
\end{equation}
\begin{equation}
\Omega_{l}=\int_{\mathbb{R}}dpM_{l}\left(p-p_{\textrm{cen}};T_{p}\right)\left|p\right\rangle \left\langle p\right|,
\end{equation}
\end{subequations} 
where $M$ is a ``mask function" modelling the detector aperture, $T_x$ and $T_p$ play the role of coarse-graining widths, and we allow extra displacement parameters $x_{\textrm{cen}}$ and $p_{\textrm{cen}}$ representing positioning degrees of freedom setting the masks'
origins. We now define $d$-dimensional periodic coarse graining (PCG) by considering the mask functions  ($j=0,\dots,d-1$) 
\begin{equation} 
\label{Eq:MaskFuncDef} 
M_j(z;T)=\left\{ \begin{array}{ccc}   1, &\; j \,s \leq z  {\rm \,(mod \, T)}  < (j+1) s \\   0, &  {\rm otherwise}  \end{array} \right. ,
\end{equation} 
as periodic square waves with spatial period $T$ and bin width $s=T/d$. The periodic functions \eqref{Eq:MaskFuncDef} define $d$ orthogonal regions covering the whole CV domain: $\sum_{k=0}^{d-1}\Pi_{k}=\mathbb{I}=\sum_{l=0}^{d-1}\Omega_{l}$. This model of coarse graining thus assigns a discrete (and finite) measurement outcome ``$j$'' to the detection of the quantum particle's CV degree of freedom ``$z$'' within the region defined by the mask function $M_j(z;T)$.
In Ref. \cite{Tasca15}, a variant of the periodic masks \eqref{Eq:MaskFuncDef} was used as analyser to test for spatial entanglement of photon pairs.

We are ready to establish the main theoretical result. If
\begin{equation}
\frac{T_{x}T_{p}}{2\pi\hbar}=\frac{d}{m},\quad m\in\mathbb{N},\quad\forall_{n=1,\ldots,d-1}\;\frac{m\,n}{d}\notin\mathbb{N},\label{MUMcondition}
\end{equation}
then the projective measurements (\ref{Proj}) with the mask function (\ref{Eq:MaskFuncDef}) fulfill (\ref{MUBs}),
thus being mutually unbiased. Since for $(l+1)d\geq m\geq ld$ with $l\in\mathbb{N}$
one finds $mn/d=nl+\left[m\;\left(\textrm{mod d}\right)\right]n/d$,
the last condition in general is concerned with $m':=m\;\left(\textrm{mod d}\right)$.  Of special significance is the case $m'=1$, since the discussed condition is valid for all dimensions $d$. On the other hand, $m'=0$ is always excluded. To gain more intuition we observe that, for instance, if $d=7$ then all values of $m'$ are allowed (obviously except $m'=0$) since $7$ is a prime number, while for $d=10$ only  $m'=1,3,7,9$ fulfill the right condition (note that $n=4,6,8$ all rule out $m'=5$). For $d=9$, we obtain $m'=1,2,4,5,7,8$, while for $d=8$ we get $m'=1,3,5,7$. Let us finally mention that  $x_{\textrm{cen}}$ and $p_{\textrm{cen}}$ play no role for unbiasedness. 
Before proceeding further, note that first part of \eqref{MUMcondition} can also be put in equivalent forms: (a) $s_xs_p=2\pi/ m d$, (b) $T_xs_p=2\pi/m$ or (c) $s_xT_p=2\pi/m$. 

To show (\ref{MUB1}), we represent \cite{Supplement} the probability associated
with $\Omega_{l}$ in terms of the position autocorrelation function \cite{ChUR}:
\begin{equation}
p_{l}\left(\Psi\right)=\frac{1}{d}+\!\!\!\!\sum_{N\in\mathbb{Z}/\{0\}}\!\!\!\!\frac{1-e^{-\frac{2\pi iN}{d}}}{2\pi iN}e^{iN\!\varphi_{l}}\!\!\!\int_{\mathbb{R}}\!\!dx\psi^{*}\!\!\left(x\right)\!\psi\left(x+N\tau_{p}\right),\label{pl}
\end{equation}
with $\psi\left(x\right)=\left\langle x\left|\Psi\right\rangle \right.\!$,
$\varphi_{l}=-2\pi l/d-p_{\textrm{cen}}\tau_{p}/\hbar$ and $\tau_{p}=2\pi\hbar/T_{p}$.
Assume first that $q_{k}\left(\Psi\right)=\delta_{k_{0}k}$ for an
arbitrary $k_{0}$. This means that $\psi\left(x\right)$ is localized
within a single periodic mask, so that the autocorrelation term, which
due to (\ref{MUMcondition}) equals $\psi^{*}\left(x\right)\psi\left(x+mNT_{x}/d\right)$,
does not vanish only when $mN/d$ is an integer. Due to the further
requirement, however, $mN/d\in\mathbb{Z}$ if and only if $N/d\in\mathbb{Z}$.
But in this special case the factor $1-e^{-\frac{2\pi iN}{d}}$ becomes
equal to $0$, so that all terms in the sum in (\ref{pl}) do vanish,
leaving the bare contribution $1/d$. The same type of derivation
applies to (\ref{MUB2}). To conclude, the set (\ref{Proj}) is mutually unbiased in a finite-dimensional manner, even though traces of all products $\Pi_k\Omega_l$ are clearly infinite.

\begin{figure}
\begin{center}
\includegraphics[width=86mm]{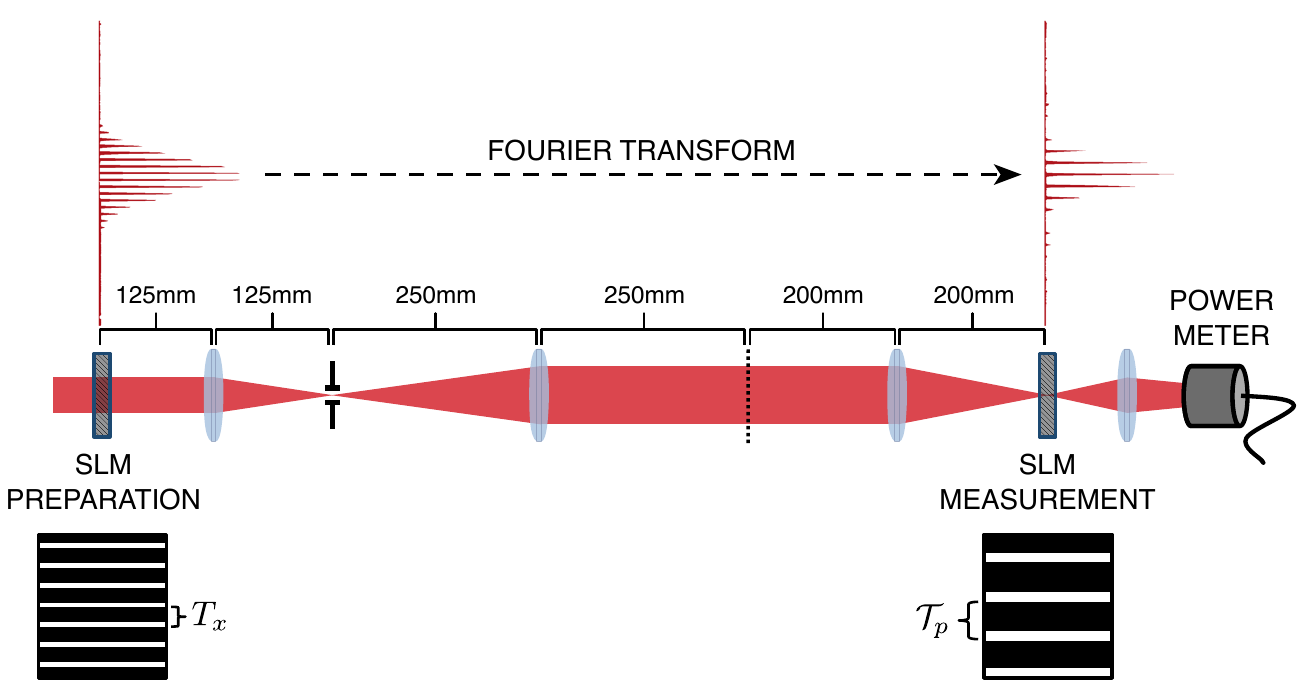}
\caption{Sketch of the experimental setup used for the demonstration of unbiased coarse-grained measurements. The transverse field distribution of a laser beam is prepared and measured using periodic spatial masks displayed on SLMs. Preparation and measurement sites are connected via optical Fourier transform. The light power transmitted through preparation and measurement spatial masks is monitored with an optical power meter.}
\label{ExpSetup}
\end{center}
\end{figure}

\paragraph{Experimental Setup.} 
The formal analogy between paraxial optics and non-relativistic quantum mechanics \cite{marcusebook} allows us to experimentally verify the condition of mutual unbiasedness \eqref{MUMcondition} using a simple optical setup implementing Fraunhofer diffraction through a multiple slit aperture. As sketched in Fig. \ref{ExpSetup}, a paraxial HeNe laser beam diffracts from preparation to measurement sites placed at the front and back focal planes of a Fourier transform lens system, respectively. At both sites, we use a spatial light modulator (SLM) to display amplitude spatial masks modelled according to our periodic coarse graining \eqref{Eq:MaskFuncDef} along the vertical direction. The illumination of the preparation SLM by the collimated laser beam with Gaussian transverse profile generates a periodically-modulated beam whose intensity distribution at the measurement SLM is that of an interference pattern produced by a periodic diffraction grating, as illustrated in Fig. \ref{ExpSetup}. The intensity distribution of this diffracted beam is then analyzed by periodic spatial masks displayed on the measurement SLM~\footnote{Our actual Lab setup uses double passage by a single SLM whose screen is divided into two sections for preparation and measurement parts of the experiment. Our device is a HOLOEYE PLUTO-NIR-015 phase-only spatial light modulator.}. 

In our experiment, the conjugate CV stand for the transverse position ($x$) and momentum ($p$) of the paraxial light field. As the transverse spatial variables at preparation and measurement sites are related via Fourier transform, the positions at the measurement SLM correspond to the transverse momentum component at the preparation SLM. Denoting $\mathcal{T}_p$ as the physical periodicity (units of length) of the spatial masks applied to the measurement SLM, we translate it to momentum domain as $T_p=\mathcal{T}_p/\alpha$, where the constant $\alpha=f_e\lambda/(2\pi)$ relates to the optical Fourier transform (we set $\hbar=1$): $f_e=100$mm is the effective focal length; $\lambda=633$nm is the light field wavelength. In terms of the physical periodicities and experimental parameters, condition \eqref{MUMcondition} reads $T_x\mathcal{T}_p=f_e \lambda d/m$.

\paragraph{Results.} 
Let us denote by $\ket{\Psi_k}=\mathcal{N}^{-1}_k \Pi_k \ket{\Psi} $ the projections of the state onto the $k$-th mask ($\mathcal{N}_k$ is a normalization constant), which by construction are eigenstates of $\Pi_k$. We perform the experiment with the building function $\psi(x)\equiv \left\langle x\left|\Psi\right\rangle \right.\!$ a Gaussian given by the transverse profile of the laser beam at the preparation SLM: $\psi(x) \propto \exp(-x^2/(4\sigma^2))$, with $\sigma=520\mu$m.

Our strategy to investigate PCG measurements is the experimental reconstruction of the probabilities $p_{kl}\equiv p_l (\Psi_k)$. The relevant distribution to evaluate unbiasedness is the conditional probability $p_{l|k}=p_{kl}/\sum_lp_{kl}$ that the outcome of PCG measurements in momentum domain is $l$, given that $\ket{\Psi_k}$ was prepared. 
As a quantifier of unbiasedness, we calculate the entropy of the distribution $p_{l|k}$:
\begin{equation}\label{Entropy}
E_k=-\sum_{l=0}^{d-1} p_{l|k} \log_2 (p_{l|k}).
\end{equation}
Hence, unbiasedness is verified whenever $E_k=\log_2 (d)$. In our setup, these outcome probabilities are obtained from the overall light power, $W_{kl}$, transmitted through the preparation and measurement spatial masks: $p_{l|k}=W_{kl}/\sum_lW_{kl}$. The transmitted light is monitored by an optical power meter (Newport 2931-C) set to output the mean value of $1000$ power measurements performed over a total sampling time of $1$s.

\begin{figure}[tbp]
\begin{center}
\includegraphics[width=86mm]{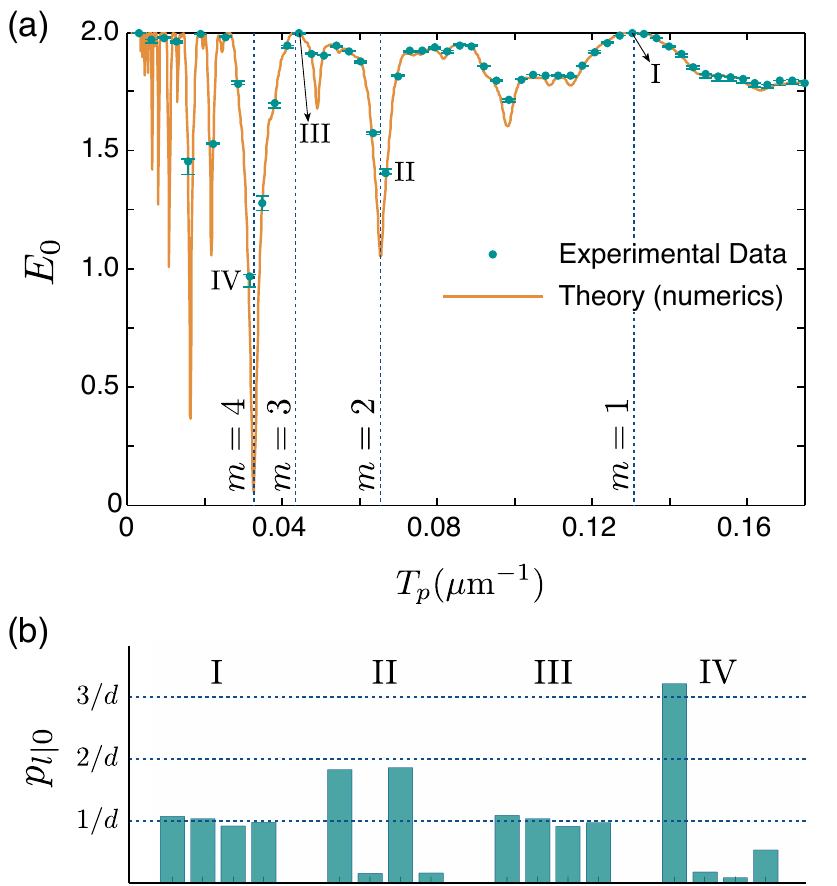}
\caption{(a) Entropy plot associated with outcomes' probabilities of PCG measurements with $d=4$ and $T_x=192\mu$m, as a function of $T_p$. (b) Examples of measured outcome  distributions $p_{l|0}$ for the selected data points shown in plot (a). Unbiasedness ($p_{l|0} \approx 1/d$) is shown for data points $\mathrm{I}$ and $\mathrm{III}$.}
\label{Scan48}
\end{center}
\end{figure}

The upper part of Fig. \ref{ExpSetup} presents an example of the prepared beam intensity distribution and its corresponding Fraunhofer diffraction pattern. For this preparation, we used the periodic spatial mask $k=0$, with bin width $s_x=48\mu$m and $d=4$, thus yielding a periodicity of $T_x=192\mu$m. Following our notation, the prepared transverse field distribution corresponds to a quantum wave-function $\psi_0(x)=\langle x| \Psi_0\rangle$. This field distribution is optically Fourier transformed and the resulting Fraunhofer diffraction pattern is subjected to PCG measurements at the measurement site. The entropy $E_0$ associated with the outcome probabilities $p_{l|0}$ is given in Fig. \ref{Scan48}-(a) as a function of the PCG periodicities in momentum domain. Experimental data are shown as turquoise points while the solid orange line represents a theoretical prediction based on numerical calculations. The momentum periodicities $T_p=2\pi d/(T_x m)$ arising from the unbiasedness condition \eqref{MUMcondition} are indicated as dashed vertical lines. It is clear that for these specific values of the periodicity the entropy assumes either a maximum (when $m$ is odd) or a minimum (when $m$ is even). In Fig. \ref{Scan48}-(b), we show the measured distributions $p_{l|0}$ associated with the data points lying closer to these periodicities. For $d=4$, the unbiasedness condition \eqref{MUMcondition} is fulfilled for $m'=1,3$, as can be evidenced from the flat probability distribution $p_{l|0}\approx 1/d$ achieved for these periodicities. 

The results presented in Fig. \ref{Scan48} illustrate unbiased PCG  measurements of dimension $4$ for the single preparation $|\Psi_0\rangle$. In order to fully demonstrate \eqref{MUB1}, we also run our experiment with complementary preparations $|\Psi_k\rangle$ ($k=0,\ldots,3$). The extreme values obtained for the entropy $E_k$ are indicated as error bars in plot \ref{Scan48}-(a). We find the values $E_k\geqslant1.9953 \pm 0.0008$ $\forall$ $k$ when using the periodicities associated with $m=1$ and $3$, thus demonstrating the full unbiasedeness relation \eqref{MUB1} between preparation and measurement outcomes for our periodic coarse graining. The presented error of $8\times10^{-4}$ is due to fluctuations of the transmitted light power over $1000$ measurements. This error is much smaller than the data points in Fig. \ref{Scan48}-(a). 
Finally, we note that the unitarity of the Fourier transformation (in mathematical terms, projection valued measures for position and momentum are unitarily equivalent \cite{Cassinelli}) and the functional dependence of condition (\ref{MUMcondition}) on the product of periodicities $T_xT_p$ (so they can be swapped), exempt the experimental verification of (\ref{MUB2}) for the demonstration of mutual unbiasedness. 

A few remarks on the resolution limitations of coarse-graining measurements are in order. The momentum periodicity $T_p$ in plot \ref{Scan48}-(a) was scanned at the resolution limit of our setup: consecutive data points relate to bin widths differing by only $8\mu$m (a single pixel of our SLM). This is the reason as to why the features shown in the theoretical prediction for smaller $T_p$ could not be demonstrated experimentally. Theoretically, Eq. \eqref{MUMcondition} provides an infinite number of possibilities for unbiasedness, but only a few are reachable in practice. The condition with $m=1$ (valid for all $d$) offers the best trade-off between experimentally attainable resolutions in the position and momentum domains. In our optical setup, this trade-off implies the physical periodicity in the measurement SLM given by $\mathcal{T}_p=f_e\lambda d/T_x$, which for $d=4$ and $T_x=192\mu$m yields $\mathcal{T}_p \approx 1319\mu$m. The focal length $f_e$ serves as a magnification parameter that can be used to adjust the momentum resolution at the cost of changing the detection range. Thus, a compromise between momentum resolution and a sensible detection range (the height of our SLM) also comes into play.

\begin{figure}[tbp]
\begin{center}
\includegraphics[width=85mm]{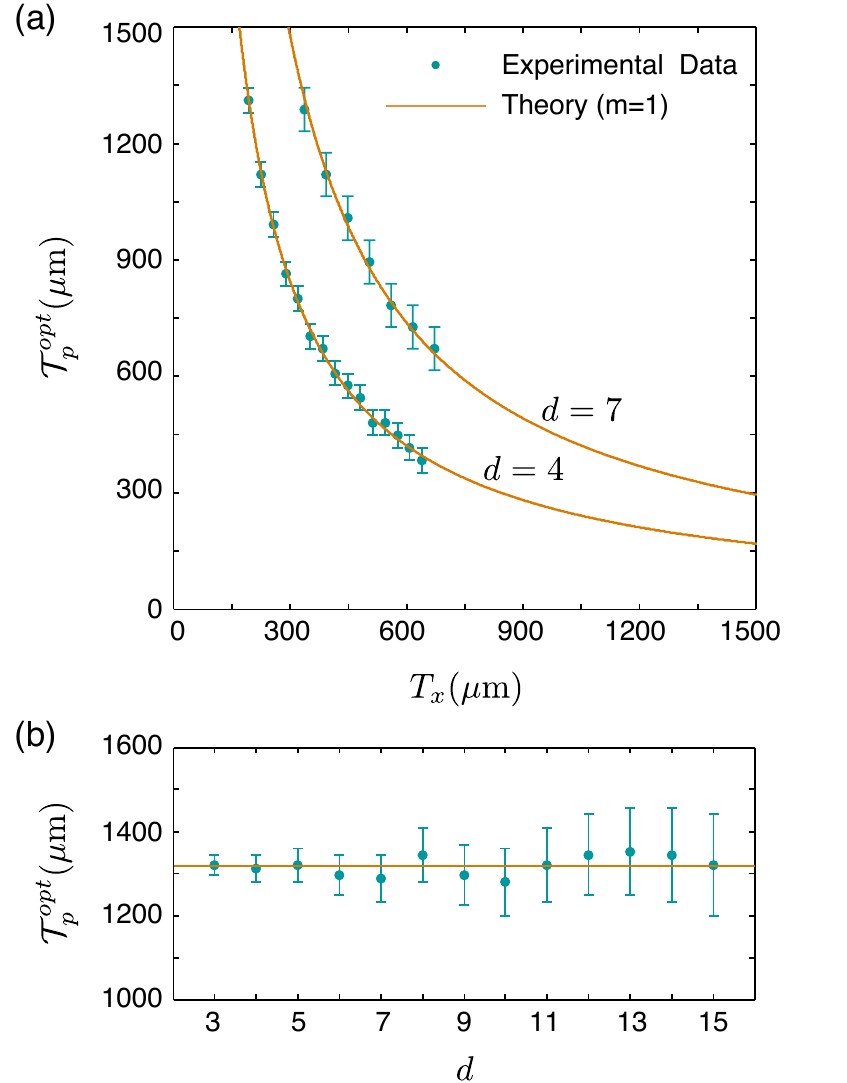}
\caption{(a) Experimental demonstration of the resolution trade-off between the pair of periodicities in position and momentum domain leading to unbiased PCG measurements with $d=4$ and $d=7.$  (b) Measured optimal periodicity in momentum domain when using mask functions with constant bin width $s_x=48\mu$m for the preparation of $d$-dimensional PCG. The theoretically predicted value is $\mathcal{T}_p^{opt}=f_e\lambda/s_x\approx 1319 \mu$m.}
\label{TpOpt}
\end{center}
\end{figure}

To illustrate the capability of our setup to implement unbiased PCG measurements, we run our system further with increased mask periodicities in the preparation SLM. For each chosen $T_x$, we look for the entropy peak associated with condition $m=1$ and experimentally determine the optimal periodicity $\mathcal{T}_p^{ opt}$ at the measurement site. These results are displayed in Fig. \ref{TpOpt}-(a) for $d=4$ and $7$. In all measurements, we obtain  optimal periodicities very close to the theoretical prediction (solid orange curve), the uncertainty in its determination (the error bars) being dictated by SLM pixel size: $ {\rm error}(\mathcal{T}_p^{ opt})=d\times 8\mu$m. For both dimensions, all obtained entropies are indicative of unbiased measurement outcomes:
$E_k\geqslant 1.997 \pm 0.001$ ($\approx \log_24$) or $E_k\geqslant 2.799 \pm 0.001$ ($\approx \log_27$).

As a final remark, the measurements  shown in Fig. \ref{TpOpt}-(b) illustrate a particular feature of \eqref{MUMcondition}. Keeping the preparation bin width at the constant value of $s_x=48\mu$m, we looked for the optimal periodicities while varying the PCG measurement dimensionality from $d=3$ up to $15$. 
In this case, the condition for the optimal periodicity is independent of the dimension parameter $d$:  $\mathcal{T}_p^{opt}=f_e\lambda/s_x \approx 1319 \mu$m.
Our measurements demonstrate close agreement with the theoretical prediction, as seen in Fig. \ref{TpOpt}-(b). We obtain $|E_k-\log_2(d)|\lesssim 0.004$ for all data points, thus demonstrating unbiased coarse-grained measurements for dimensionality up to $15$. 

\paragraph{Discussion.}
 We have shown how one can recover the condition of mutual unbiasedness in coarse-grained measurements of continuous variable systems.  Periodic ``mask'' functions were used to define projective measurements in position and momentum variables, and these were shown to be mutually unbiased for particular combinations of periodicities. What other types of mask functions satisfy the MUB condition is an open question. Non-trivial coarse-graining structures have already been considered in the context of Bell inequalities violations \cite{Wenger03,Cavalcanti11,Quintino12} and quantum cryptography \cite{Walborn06} with CV systems. Nonetheless, a formal demonstration of mutual unbiasedness was still missing in the general framework of CV measurements. We believe our results may provide a method to formally connect continuous and discrete variable quantum mechanics.     

\begin{acknowledgements}
D.S.T. and P.S. thank A. Z. Khoury for discussions and access to Lab facilities.
The authors acknowledge financial support from the Brazilian funding agencies CNPq, CAPES and FAPERJ, as well as from the Brazilian National Institute of Science and Technology for Quantum Information. P.S. acknowledges financial support by grant number 233-2015 from CONCYTEC-FONDECYT, Peru. \L .R. acknowledges financial support by grant number 2014/13/D/ST2/01886 of the National Science Center, Poland. 
\end{acknowledgements}

\onecolumngrid
\clearpage
\begin{center}
\textbf{\large Supplemental Material for \textit{Mutual Unbiasedness in Coarse-grained Continuous Variables}} \\ \vspace{0.2in}
Daniel S. Tasca, Piero S\'{a}nchez, Stephen P. Walborn, and \L ukasz Rudnicki \\ \vspace{0.4in}
\end{center}

\setcounter{equation}{0}
\setcounter{figure}{0}
\setcounter{table}{0}
\setcounter{page}{1}
\setcounter{section}{0}
\makeatletter
\renewcommand{\theequation}{S\arabic{equation}}
\renewcommand{\thefigure}{S\arabic{figure}}

The aim of this Supplemental Material is to express $p_{l}\left(\Psi\right)=\left\langle \Psi\right|\Omega_{l}\left|\Psi\right\rangle $
defined by
\begin{equation}
\Omega_{l}=\int_{\mathbb{R}}dpM_{l}\left(p-p_{\textrm{cen}};T_{p}\right)\left|p\right\rangle \left\langle p\right|,
\end{equation}
with
\begin{equation} \label{Eq:MaskFuncDefsup} M_k(z;T)=\left\{ \begin{array}{ccc}   1, &\; k \,s \leq z  {\rm \,(mod \, T)}  < (k+1) s \\   0, &  {\rm otherwise}  \end{array} \right. ,\qquad k=0,\dots,d-1,
\end{equation} in terms of the position wave-function $\psi\left(x\right)=\left\langle x\left|\Psi\right\rangle \right.\!$.
Note that, by definition
\begin{equation}
p_{l}\left(\Psi\right)=\int_{\mathbb{R}}dpM_{l}\left(p-p_{\textrm{cen}};T_{p}\right)\tilde{\rho}\left(p\right),
\end{equation}
where $\tilde{\rho}\left(p\right)=|\tilde{\psi}\left(p\right)|^{2}$
and $\tilde{\psi}\left(p\right)=\left\langle p\left|\Psi\right\rangle \right.\!$. 

To achieve the desired goal we first represent the periodic mask function in terms of its Fourier decomposition
\begin{equation}
M_{k}\left(z;T\right)=\sum_{N\in\mathbb{Z}}\frac{1-e^{-\frac{2\pi iN}{d}}}{2\pi iN}e^{-\frac{2\pi iN}{d}k}e^{\frac{2\pi iN}{T}z}\equiv\frac{1}{d}+\!\!\sum_{N\in\mathbb{Z}/\{0\}}\!\!\frac{1-e^{-\frac{2\pi iN}{d}}}{2\pi iN}e^{-\frac{2\pi iN}{d}k}e^{\frac{2\pi iN}{T}z}.
\end{equation}
Clearly
\begin{equation}
p_{l}\left(\Psi\right)=\frac{1}{d}+\!\!\sum_{N\in\mathbb{Z}/\{0\}}\!\!\frac{1-e^{-\frac{2\pi iN}{d}}}{2\pi iN}e^{iN\!\phi_{l}}\int_{\mathbb{R}}dpe^{\frac{iN\tau_{p}}{\hbar}p}\tilde{\rho}\left(p\right),
\end{equation}
with $\phi_{l}=-2\pi l/d-p_{\textrm{cen}}\tau_{p}/\hbar$ and $\tau_{p}=2\pi\hbar/T_{p}$.
The momentum integral gives the characteristic function of the momentum
probability distribution, $\tilde{\Phi}\left(N\tau_{p}/\hbar\right)$,
defined as
\begin{equation}
\tilde{\Phi}\left(\lambda\right)=\int_{\mathbb{R}}\!\!dp\,e^{i\lambda p}\tilde{\rho}\left(p\right).
\end{equation}

The autocorrelation form of the characteristic function reads \cite{ChURsup}:
\begin{equation}
\tilde{\Phi}\left(\lambda\right)=\int_{\mathbb{R}}\!\!dx\psi^{*}\!\left(x\right)\!\psi\left(x+\hbar\lambda\right).\label{autocorrelation}
\end{equation}
Note that Eq. 3 from \cite{ChUR} deals with the characteristic function
of position probability distribution, so that the shift $\hbar\lambda$
appears with the minus sign. The formula (\ref{autocorrelation})
is a simple consequence of the Fourier transformation between wave
functions in both domains. As a final result we obtain the desired
expression
\begin{equation}
p_{l}\left(\Psi\right)=\frac{1}{d}+\!\!\sum_{N\in\mathbb{Z}/\{0\}}\!\!\frac{1-e^{-\frac{2\pi iN}{d}}}{2\pi iN}e^{iN\!\phi_{l}}\!\int_{\mathbb{R}}\!\!dx\psi^{*}\!\left(x\right)\!\psi\left(x+N\tau_{p}\right).\label{pl}
\end{equation}

\end{document}